\documentclass[12pt,titlepage]{article}

\usepackage{latexsym}
\usepackage{amsmath}
\usepackage{amssymb}
\usepackage{mathrsfs}
\usepackage{dsfont}
\usepackage{slashed}
\usepackage{stmaryrd}
\usepackage{youngtab}
\usepackage[dvips]{graphicx}
\usepackage{pdfsync}
\usepackage[bookmarks=false,colorlinks=true,linkcolor=blue,citecolor=blue,urlcolor=green,linktoc=page]{hyperref}
\usepackage{accents}

\DeclareSymbolFont{txfont}{OT1}{txr}{m}{it}
\DeclareMathSymbol{\varw}{\mathalpha}{txfont}{119}
\DeclareMathSymbol{\varg}{\mathalpha}{txfont}{103}
\DeclareMathSymbol{\vari}{\mathalpha}{txfont}{105}
\DeclareMathSymbol{\imi}{\mathalpha}{txfont}{105}

\newcommand{\hal}{{\textstyle\frac{1}{2}}}

\newcommand{\ca}[1]{{\mathcal{#1}}}

\newcommand{\unit}{\mathds{1}}

\newcommand{\del}{\partial}
\newcommand{\vep}{\varepsilon}

\newcommand{\w}{{\scriptstyle{\wedge}}\,}
\newcommand{\ww}{\mathfrak}

\newcommand{\be}{\begin{equation}}
\newcommand{\ee}{\end{equation}}
\newcommand{\ben}{\begin{displaymath}}
\newcommand{\een}{\end{displaymath}}
\newcommand{\bea}{\begin{eqnarray}}
\newcommand{\eea}{\end{eqnarray}}
\newcommand{\nn}{\nonumber}

\newcommand{\bean}{\begin{eqnarray*}}
\newcommand{\eean}{\end{eqnarray*}}
\newcommand{\beqs}{\begin{eqnarray}}
\newcommand{\eeqs}{\end{eqnarray}}
\newcommand{\beal}{\begin{align}}

\newcommand{\cm}{{\ww{m}}}

\makeatletter
\@addtoreset{equation}{section}
\makeatother

\begin{document}


\begin{titlepage}

 \begin{flushright}\small
 MIFPA-12-45\\
 YITP-SB-12-46\\
   December 2012 
 \end{flushright}

\vspace{.8cm}

 \begin{flushleft}

  {\LARGE \sffamily \bfseries Six-dimensional superconformal couplings of \\[.7ex]
    non-abelian tensor and hypermultiplets}
  \\[0.8cm]
  \rule{\linewidth}{0.5mm}
  \\[0.8cm]
  { \sffamily \bfseries  Henning Samtleben$^{a}$, Ergin Sezgin$^b$, Robert Wimmer$^{a,c}$}
  \\[1cm]

  \begin{minipage}{\textwidth}
    \begin{flushleft}
    {\small\it\raggedright
        $^a$Universit\'e de Lyon, Laboratoire de Physique, UMR 5672, CNRS, \\
         \'Ecole Normale Sup\'erieure de Lyon,\\
         46, all\'ee d'Italie, F-69364 Lyon cedex 07, France\\
         \vskip 3mm
         \noindent
         $^b$\,\it\raggedright  George P. and Cynthia W. Mitchell Institute \\for Fundamental
                Physics and Astronomy \\
                Texas A\&M University, College Station, TX 77843-4242, USA\\
                \vskip 3mm
         \noindent
        $^c$\,\it\raggedright C.N. Yang Institute for Theoretical Physics\\
       State University of New York\\
       Stony Brook, NY 11794-3840, USA.} \\[0.5cm]
    {{\it Email}: \tt henning.samtleben@ens-lyon.fr, sezgin@physics.tamu.edu,
                      wimmer@insti.physics.sunysb.edu}

     \end{flushleft}
  \end{minipage}

\vspace{2cm}

\parbox\textwidth{A{\small BSTRACT}: 
We construct six-dimensional superconformal models with non-abelian tensor and hypermultiplets.
They describe the field content of $(2,0)$ theories, coupled to $(1,0)$ vector multiplets.
The latter are part of the non-abelian gauge structure that also includes 
non-dynamical three-  and four-forms.
The hypermultiplets are described by gauged nonlinear sigma models with a hyper-K\"ahler 
cone target space. 
We also address the question of constraints in these models and show that their resolution
requires the inclusion of abelian factors. These provide couplings that were previously
considered for anomaly cancellations with abelian tensor multiplets and resulted in the selection
of  $ADE$ gauge groups.
}

\vfill

  \end{flushleft}
\end{titlepage}

\noindent\rule\textwidth{.1pt}		
      \vskip 2em
      \tableofcontents
      \vskip 2em
      \noindent\rule\textwidth{.1pt}
      \vskip 2em

\section{Introduction}

For a long time it was believed that four is the maximal dimension in which 
conformal field theories can exist. However, string/M-theory suggest that the
group theoretical bound  on superconformal theories of six dimensions \cite{Nahm:1977tg}
is actually saturated \cite{Witten:1995ex}. 
The understanding of effective (2,0) theories for multiple M5-branes is one of the 
pressing questions of M-theory in this context. They are classified by $ADE$ groups,
but an explicit understanding still seems far off.

One of the problems is of pure geometrical origin and independent
of any underlying dynamics or supersymmetry and addresses the question of 
non-abelian tensor (two-form) gauge fields. For example, various no-go theorems
exclude the non-abelian extension of the abelian tensor gauge symmetry \cite{Bekaert:2000qx}.
In~\cite{Samtleben:2011fj} this problem was encompassed in the context of a tensor hierarchy 
\cite{deWit:2005hv,deWit:2008ta} by introducing additional form-degrees of freedom, 
in particular an ordinary gauge field and non-propagating three- and (optionally) four-form gauge fields. 
This structure shows similarity with concepts of higher gauge theories, Q structures, and non-abelian gerbes
\cite{Kotov:2007nr,Baez:2010ya,Kotov:2010wr,Fiorenza:2012tb,Saemann:2012uq,Palmer:2012ya}, extended to higher degree forms. 
A very particular realization of this gauge symmetry was given in~\cite{Chu:2011fd}.

The other problem is that the supposed $(2,0)$ theory of multiple M$5$-branes is intrinsically strongly
coupled, i.e.\ it has no free parameter for a weak coupling expansion which would make the
existence of a Lagrangian description plausible. This problem is analogous to 
the situation of M$2$-branes. Also in that case, and for the same reason, it was believed that a Lagrangian 
description does not exist. Nevertheless, a single maximally supersymmetric three-dimensional 
CFT (BLG-model) \cite{Bagger:2007jr,Gustavsson:2007vu} 
and a more general class  with less supersymmetry  (ABJM-models) \cite{Aharony:2008ug} have been found.
The decisive observation in the latter case is that by placing the M$2$-branes at an orbifold singularity 
instead of placing them in flat space one gains a dimensionless parameter which allows for a weak coupling 
limit and thus makes a Lagrangian description possible.  The resulting CFT's have the same field content as a 
maximally, i.e.\ ${\cal N}=8$ supersymmetric theory but realize only ${\cal N}=6$ supersymmetry. From the field 
theory point of view this means that the reduced supersymmetry is less restrictive and therefore allows for a 
Lagrangian formulation.
\\
\noindent

In a similar spirit, in~\cite{Samtleben:2011fj,Samtleben:2012mi} we constructed 
interacting $(1,0)$ superconformal models for 
non-abelian tensor and vector multiplets.
The BPS sector of these models has been analyzed in \cite{Akyol:2012cq}.
Also in six dimensions, $\ca{N}= (1,0)$ supersymmetry is not as restrictive 
as to prevent any nontrivial local dynamics (as in the maximal $(2,0)$ case), 
but  strong enough to essentially determine the dynamics.
As a further step towards the sought-after (2,0) models, in this paper we complete the 
$(1,0)$ tensor multiplet interactions
to the full field content of the $(2,0)$ theories by coupling the 
non-abelian tensor/vector models of  
\cite{Samtleben:2011fj,Samtleben:2012mi} to superconformal hypermultiplets.

Superconformal hypermultiplets are described by a gauged non-linear sigma model \cite{Sierra:1983uh}. 
Conformal invariance requires the target space to be a hyper-K\"ahler cone (HKC) 
\cite{deWit:1999fp, deWit:2001bk}. 
The possible gauge groups are subgroups of the isometry group of the underlying quaternionic 
K\"ahler manifold (QK). Supersymmetry 
requires that the vector multiplets that gauge these isometries are embedded in a particular way into the vector
multiplets of the superconformal vector/tensor system. There is no direct coupling between the 
tensor and hypermultiplets prior to eliminating the auxiliary fields, 
but the vector multiplets form the `glue' between this two multiplets
in a non-trivial way, beyond simple minimal gauge couplings.   
Though the restriction of the gauge group by the QK isometry group suggests some selection 
mechanism there remains a large freedom in the construction. Even the restriction to compact quaternionic 
K\"ahler manifolds provides for all classical groups a corresponding manifold, the Wolf spaces \cite{Wolf:1965}, 
and an associated HKC. We discuss in detail the case of the flat HKC and show that one can embed arbitrary 
matrix representations of semi-simple groups including abelian factors in the corresponding isometry group which 
 in this case is $Sp(n)/\mathbb{Z}_2$. It is however not surprising that pure classical 
considerations do not lead to the 
selection of $ADE$ groups since these are determined by anomaly cancellation conditions and is thus an essential
quantum effect. For abelian tensor multiplets this was discussed in~\cite{Blum:1997mm}.

From the different types of superconformal tensor/vector models of~\cite{Samtleben:2012mi} the coupling
to the hypermultiplets selects those whose field equations can be integrated to an action 
(modulo the known subtleties related to the description of self-dual tensor fields).
Upon extending the system to include non-propagating four-form potentials, 
the dynamics may equivalently be expressed by a set of non-abelian first-order duality equations.
This description appears rather natural with the coupling to hypermultiplets. 
The resulting supermultiplet structure is rather intriguing. While the gauge structure
based on the three-form leads to an on-shell supermultiplet that mixes the 
tensor and vector multiplet with the non-dynamical three-form, the inclusion of the four-form 
also intertwines the hypermultiplets with the previous ones. 
Even more intriguing is the observation, that in the supersymmetry 
transformation of the four-form the tensor and hypermultiplet contributions  
combine in a manifestly $(2,0)$ way.

Another question that we address in this paper is the elimination of the vector multiplet auxiliary fields.
As has been mentioned, the vector multiplet forms the `glue' between the tensor and hypermultiplet. 
A particular important coupling comes through the auxiliary field, which is described by the 
algebraic field equation,
\begin{equation}
  \label{eq:1}
  d_{Irs}\,\left(\, Y^s_{ij} \,\phi^I - 2 \bar\lambda^s_{(i} \chi_{j)}^I \,\right) 
  -\frac{1}{2\lambda}\, \theta_r{}^{\ww{m}}\,\mu^{ij}_{\ww{m}} =0\, ,
\end{equation}
where $Y$ is the auxiliary field and $\lambda^s_i$ the gaugino of the vector multiplet, while
$\phi,\chi$ belong to the tensor multiplet. The moment maps $\mu$ of the hypermultiplet Lagrangian
couple with a free dimensionless coupling constant $\lambda$. The other two objects are invariant tensors
of the gauge group. This equation in fact contains the full information about the vector multiplet dynamics,
the rest is fixed by supersymmetry. Generically this equation implies 
constraints on the elementary fields \cite{Samtleben:2011fj,Samtleben:2012mi}. The inclusion of the 
hypermultiplets does not alter this observation. 
We show that one has to include abelian factors, or tensor multiplet singlets, in order to avoid
constraints for the elementary fields. In that case one finds a unique solution for the auxiliary fields $Y$.
In contrast to the standard YM-hyper couplings of~\cite{Sierra:1983uh} however, 
eliminating the auxiliary fields does not generate a bosonic potential but only
quadratic and quartic fermionic interactions.

The resulting moduli space for the scalars is thus not constrained by any potential. In particular the
VEV for the tensor scalar fields can in principle take any value. However, we find certain couplings 
between the tensor scalars and the vector multiplet such that the VEV $\langle \phi\rangle$ acts as the 
inverse (square) of the  Yang-Mills coupling constant. 
Therefore at the conformal point $\langle \phi\rangle =0$ 
the theory is no longer perturbatively defined. Also the auxiliary field equation (\ref{eq:1}) is degenerate in 
this point. It implies constraints on the hypermultiplets and thus changes the target space geometry. 
This is of course the indication of the well known, but not so well understood,  phenomenon of 
the tensionless string phase transition \cite{Seiberg:1996vs,Duff:1996cf}. Consequently the models are 
well defined only on the Coulomb branch, where the conformal symmetry is spontaneously broken. 
However, since the breaking is spontaneous the original conformal symmetry might be still useful 
in the perturbative regime through the broken Ward identities, although they might look very complicated. 
The literature mentions evidence that this conformal fixed point might be described by a local 
field theory (see e.g. \cite{Blum:1997mm}). With the given set of degrees of freedom that we considered here, 
which comprises the full field content of $(2,0)$ tensor multiplets, 
superconformal symmetry predicts essentially unique models where this phenomenon can be at most seen as a 
highly non-perturbative effect.

A last comment regarding $ADE$ classification:  
$\ca{N}=(1,0)$ theories are chiral and therefore anomaly cancellation is an important ingredient. 
As mentioned above, the anomaly cancellation conditions of \cite{Blum:1997mm} lead to a selection 
of $ADE$ gauge groups, though the tensors are abelian in that case. 
For the models presented here, we have to postpone such a discussion 
until a full quantum treatment is available. 
However, we make the following observation: The necessity to include
abelian factors to avoid constraints in (\ref{eq:1}) also provides the structure for the particular couplings
that were considered in \cite{Blum:1997mm},
see equation~(\ref{c1}) below. These couplings, that are an essential input for the 
anomaly cancellation conditions, are thus naturally present in our theories. 
Let us finally mention other approaches towards the field 
equations~\cite{Lambert:2011gb,Saemann:2012uq,Bonetti:2012st}
and amplitudes~\cite{Czech:2011dk} of the (2,0) equations
whose relation to the presented construction will be interesting to understand.

The paper is organized as follows: In section 2 we discuss the geometrical background for the 
superconformal hypermultiplets. In section 3 we describe the Lagrangians for the hyper- and the tensor/vector
system, respectively, and the embedding of the hypermultiplet gauging into the latter. 
We also discuss the `first-order' description, 
which includes the non-dynamical four-form and its supermultiplet structure. 
In section 4 we discuss the elimination of the auxiliary fields and the necessity of 
abelian factors and resulting interactions.

\section{Geometrical setting}

The target space of rigid supersymmetric sigma models with eight supercharges has to be
a hyper-K\"ahler (HK) manifold\footnote{If one considers only equations of
motions no metric $g$ is needed and the requirements for susy are less restrictive
\cite{Rosseel:2004fa,Bergshoeff:2004nf}. However, we will always assume the existence
of an action.} $(\,\ca{M}_{4n}, g, \vec J\,)$ \cite{AlvarezGaume:1981hm, Sierra:1983fj}.
Hence, $\ca{M}_{4n}$ is
a (real) $4n$-dimensional, with local coordinates $q^{\alpha = 1,\ldots,4n}$, and provides an
$Sp(1)$-triplet of covariant constant complex structures $\vec J$ and a metric $g$ which is
hermitian w.r.t.\ all of them. This hyper-complex structure forms a quaternionic algebra
and defines a triplet of hyper-K\"ahler forms $\vec\omega$:
\begin{equation}
  \label{hk}
  J^{\mathsf{i}}J^{\mathsf{j}} = -\, \delta^{\mathsf{i}\mathsf{j}} +
   \vep^{\mathsf{i}\mathsf{j}\mathsf{k}} J^{\mathsf{k}}\ ,  \quad \nabla_\alpha \vec J =0 \ , \quad
   \vec\omega_{\alpha\beta}:=g_{\alpha\gamma}\vec{J}^{\ \gamma}{}_\beta,
\end{equation}
where $\nabla_{\alpha}$ is the Levi-Civita connection. This actually implies the existence of an whole
$S^2\simeq\mathds{C}P^1$ of complex structures $\{\mathcal I =\vec a\cdot\vec J\mid \vec{a}^2 = 1\}$, and that
the K\"ahler forms are closed, $d\,\vec\omega = 0$. The latter, or the requirement of reduced holonomy $\mathrm{Hol}(g)\subseteq Sp(n)$ may be taken as an equivalent definition of a HK manifold.
In the following we pick $J^3$ as the particular complex structure to define complex
coordinates\footnote{The real coordinates we define then as
$[q^\alpha] =\tfrac{1}{\sqrt{2}} [(z^a + \bar z^{\bar a}), -i(z^a - \bar z^{\bar a})]^t$\,.}
$z^a, \bar z^{\bar a}=(z^a)^*$ such that $J^3$ is diagonal and the fundamental holomorphic
$(2,0)$ form is given by:
\begin{equation}
  \label{hk2}
  J^{3\, a}{}_b = i\,\delta^a{}_b\ , \quad
  J^{3\, \bar a}{}_{\bar b} = -i\,\delta^{\bar a{}}{}_{\bar b}\ , \quad
  \omega^{(+)}:=\tfrac{1}{2}\,( \, \omega^1 + i\, \omega^2)\ .
\end{equation}
The holomorphic $(2,0)$ form $\omega^{(+)}$ implies that a HK manifold is also a
holomorphic symplectic manifold. Furthermore $(\omega^{(+)})^n$ defines a
nowhere vanishing section of the canonical bundle and thus $\ca{M}_{4n}$ is Ricci-flat.
\\

\noindent
{\bf{Hyper K\"ahler Cones.}}
For a sigma model with the given amount of supersymmetry
to be conformal the HK target space $\ca{M}_{4n}$ has to
be of special type, namely a hyper-K\"ahler-cone (HKC) \cite{deWit:1999fp}. These spaces are
characterized by the existence of a homothetic Killing vector field,
\begin{equation}
  \label{hKV}
  \nabla_\alpha\, \chi^\beta = \delta^\beta{}_\alpha\ .
\end{equation}
This implies for the Lie derivative $\mathscr{L}_{\chi}\,g_{\alpha\beta} = 2\, g_{\alpha\beta}$ and
thus $\chi^\alpha$ generates dilatations\footnote{We define the homothetic Killing vector with unit normalization
(\ref{hKV}). The actual dilatation generator will be defined with a constant factor, appropriate
for the six-dimensional world volume.}
on $\ca{M}_{4n}$, see also \cite{Sezgin:1994th}. The homothetic Killing vector defines i.) a
hyper-K\"ahler potential $\chi(q): \ca{M}_{4n} \rightarrow \mathds{R}$, i.e.\ a single
K\"ahler potential for all complex structures
$\cal{I}$ \cite{swann}, and ii.) the Killing vectors
$\vec{k}^\alpha$ which generate the $Sp(1)$ isometry that acts on the complex structures and becomes
the R-symmetry of the supersymmetry algebra:
\begin{equation}
  \label{hkp}
  \chi(q):= \tfrac{1}{2}\, g_{\alpha\beta}\chi^\alpha\chi^\beta \quad \Leftrightarrow \quad
   g_{\alpha\beta} = \nabla_\alpha\del_\beta \chi(q)\ , \quad \vec{k}^\alpha :=\tfrac{1}{2}\, \vec{J}^\alpha{}_\beta\chi^\beta\, ,
\end{equation}
from which follows
$\ca{L}_{\vec\alpha\vec k}\, J^{\mathsf{j}} = -\alpha^{\mathsf{i}}\,\vep^{\mathsf {ijk}}J^{\mathsf{k}}
=\vec\alpha\,\vec{t}^{\ \mathsf{j}}{}_{\mathsf{k}}\, J^{\mathsf{k}}$.
The equation for the metric can be considered as an equivalent definition of a HKC, where
$\chi^\alpha$ is then  obtained as $\chi^\alpha = g^{\alpha\beta}\del_\beta\chi(q)$, and it imposes rather strong conditions.
Not only are the non-vanishing components of the metric as usual derived from a
K\"ahler potential (in complex coordinates $g_{a\bar b} = \del_a \del_{\bar b}\chi$),
but it also implies that $\nabla_a\del_b\chi = 0$.

The superconformal case is closely related to the situation with local supersymmetry and the
coupling to supergravity. The latter requires that the
target space is a \mbox{${4(n-1)}$} dimensional
quaternionic K\"ahler manifold $\ca{Q}_{4(n-1)}$ \cite{Bagger:1983tt}.
There exists a one-to-one
correspondence between HKC's $\ca{M}_{4n}$ and  quaternionic K\"ahler manifolds  $\ca{Q}_{4(n-1)}$
via  the ``superconformal quotient'' \cite{deWit:1999fp, deWit:2001dj} and
the Swann bundle $\ca{M}_{4n}\rightarrow \ca{Q}_{4(n-1)}$ \cite{swann}.
The coupling to supergravity gauges the above described $Sp(1)$ isometry. 
This is in contrast with the rigid superconformal
case with global $Sp(1)$ R-symmetry, that we consider here. However, we are interested in the
construction of gauged superconformal models and the possible gauge groups are given by
those isometries of the HKC $\ca{M}_{4n}$ that can be gauged while preserving
super- and conformal symmetry.  These are the triholomorphic isometries
which commute with the $Sp(1)$ isometries (\ref{hkp}) and the dilatations (\ref{hKV}).
As a matter of fact, these are the isometries of the underlying quaternionic K\"ahler
manifold $\ca{Q}_{4(n-1)}$  \cite{deWit:2001dj, deWit:2001bk}.
Thus $\mathrm{Iso}(\ca{Q}_{4(n-1)})$ describes the  possible gauge groups of
the superconformal sigma models with the corresponding  HKC target space
$\ca{M}_{4n} \rightarrow \ca{Q}_{4(n-1)}$.

It is conjectured \cite{LebrunSalamon} that the Wolf spaces \cite{Wolf:1965} are
all possible (positive curvature) compact quaternionic manifolds
$\ca{Q}_{4(n-1)}$. These are symmetric spaces
and there exists one for each simple Lie group. The quaternionic projective space,
\begin{equation}
  \label{Qflat}
  \ca{Q}_{4(n-1)} = \mathds{H}P^{n-1} = \frac{Sp(n)}{Sp(1)\times Sp(n-1)}\ ,
\end{equation}
whose HKC is the flat space $\mathds{R}^{4n}$, will be of particular interest for us. The isometry
group in this case is  $Sp(n)/\mathbb{Z}_2$ \cite{Shankar} and thus we can realize representations
of gauge groups that can be embedded in in $Sp(n)/\mathbb{Z}_2$.
For the rest of this paper, we will restrict to discussing gaugings on the level of the Lie algebra, i.e.\ for $Sp(n)$.\\

\noindent
{\bf{Triholomorphic isometries, moment maps.}}
As has been indicated, the isometries that can be
gauged in accordance with superconformal symmetry have to be i.) triholomorphic and commute
with the $Sp(1)$ isometries in order to preserve supersymmetry \cite{Sierra:1983fj, Hull:1985pq},
and ii.) commute with the homothetic Killing vector field in order to preserve conformal invariance
\cite{deWit:2001dj}. Therefore the corresponding Killing vector fields $X_{(\hat{\ww{m}})}$ are
defined by the conditions,
\begin{equation}
  \label{3hol}
  \mathscr{L}_{X_{(\hat{\ww{m}})}}\, g_{\alpha\beta} = 0\ ,\quad
  \mathscr{L}_{X_{(\hat{\ww{m}})}}\, \vec{\omega}_{\alpha\beta} = 0\ , \quad
   [\, X_{(\hat{\ww{m}})},\vec k\, ]   = [\, X_{(\hat{\ww{m}})}, k_D\,] = 0\, ,
\end{equation}
where we have introduced the properly normalized dilatation Killing vector
$k_D^\alpha = \varw_q\, \chi^A$,
with $\varw_q=2$ for a six-dimensional world volume. Given that the K\"ahler forms $\vec\omega$ are
closed the triholomorphicity condition reduces to
$\mathscr{L}_{X_{(\hat{\ww{m}})}}\, \vec{\omega} =
d (i_{X_{(\hat{\ww{m}})}}\, \vec\omega)=0$.
This in turn implies the existence  of  $Sp(1)$-triplets of (local) moment maps
(Killing potentials)
$\vec\mu_{(\hat{\ww{m}})}$,
\begin{equation}
  \label{mmap}
  d\,\vec\mu_{(\hat{\ww{m}})} : = -i_{X_{(\hat{\ww{m}})}}\vec\omega
     \quad\Rightarrow\quad \del_\alpha\vec\mu_{(\hat{\ww{m}})}
             = \vec\omega_{\alpha\beta}X^\beta_{(\hat{\ww{m}})}\, ,
\end{equation}
which reflects the mentioned symplectic structure of the HK manifold.
The moment maps $\vec \mu_{(\hat{\ww{m}})}$ will define the potential-coupling to the
vector multiplet, that one obtains after gauging the isometries. Therefore the global
existence of the moment maps is a necessary (and sufficient \cite{bagger}) condition
to gauge the associated isometries. The moment maps (\ref{mmap}) are defined only up to
constants, which however are eventually fixed by the requirement of conformal symmetry,
see (\ref{ge11}).
\\

\noindent
{\bf{Gauging.}} The gauging of isometries, especially in the given context, has been considered
in \cite{Hull:1985pq,Sierra:1983uh,deWit:1999fp,deWit:2001bk}. The triholomorphic
Killing vector fields $X_{(\hat{\ww{m}})}$
generate the isometry group $\hat G=\mathrm{Iso}(\, \ca{Q}_{4(n-1)}\,)$ of the quaternionic
K\"ahler manifold associated to the HKC $\ca{M}_{4n}$. Generically we will gauge a
subgroup $G\subseteq \hat G$, generated by the subset of triholomorphic isometries
$\{ X_{(\ww{m})}\,\} \subseteq \{ X_{(\hat{\ww{m}})}\,\}$ that satisfy,
\begin{equation}
  \label{lie}
  [\, X_{(\ww{m})},\, X_{(\ww{n})}\,] =
       -f_{\ww{m}\ww{n}}{}^{\ww{p}} X_{(\ww{p})}\,.
\end{equation}
The $f_{\ww{m}\ww{n}}{}^{\ww{p}}$ are the structure constants of the associated
Lie algebra $\mathfrak{g}:=\mathrm{Lie}(G)$.

The target space coordinates $q^\alpha$ will eventually depend on the
six-dimensional world volume coordinates $x^\mu$. In the process of gauging the
isometry transformations are made local w.r.t.\ the world volume:
\begin{equation}
  \label{lociso}
   \delta_\Lambda\, q^\alpha(x):= \Lambda^{\ww{m}}(x)\, X_{(\ww{m})}^\alpha\, .
\end{equation}
Correspondingly one introduces gauge fields on the world-volume and covariant derivatives,
\begin{equation}
  \label{covder}
  D_\mu q^\alpha = \del_\mu q^\alpha -  A^{\ww{m}}_\mu\,  X^\alpha_{(\ww{m})}\ , \quad
  \delta_\Lambda\, A^{\ww{m}}_\mu =
   \del_\mu\Lambda^{\ww{m}}  -  A^{\ww{n}}_\mu\,\Lambda^{\ww{p}}\, f_{\ww{n}\ww{p}}{}^{\ww{m}}\, ,
\end{equation}
which results in the covariant transformation law,
\begin{equation}
  \label{dcov}
  \delta_\Lambda\, (D_\mu q^\alpha) =  \Lambda^{\ww{m}}\, (T_{\ww{m}}){}^\alpha{}_\beta\, D_\mu q^\beta \quad
   \textrm{with}\quad
   (T_{\ww{m}}){}^\alpha{}_\beta = \del_\beta X_{(\ww{m})}^\alpha\, .
\end{equation}

We finally note that in the case that $G\subset \hat G$ is a normal subgroup the original global
isometry group is preserved by the gauging and one has in addition to the
gauge invariance $G$ the global symmetry $H=\hat G/G$. However, in the generic case the original
symmetry $\hat G$ is broken down to $G$ by the gauging \cite{Sierra:1983uh}.

\paragraph{Flat target space, $Sp(n)$ isometries.}
We conclude this section with an explicit discussion for the case that the
target space is flat $\mathbb{R}^{4n}$ which will be the case of particular interest
in the following. Nevertheless, all of the subsequent constructions 
apply to general HKC target spaces.
For flat target space, all connection coefficients
can be set to zero and also the curvature tensor in (\ref{La}) vanishes.
The basic HK structures take in complex coordinates the simple form,
\begin{equation}
  \label{fHK}
 \chi = \sum_{a=1}^{2n} |z^a|^2\ , \quad  \omega^3 = -i\, dz^a\w dz^{\bar{a}}\ ,
  \quad \omega^{(+)} = \tfrac{1}{2}\, \Omega_{ab}\, dz^a\w dz^b\, ,
\end{equation}
with the $Sp(n)$ invariant symplectic form $\Omega_{ab}$ ($a,b=1,\ldots, 2n$),
while the homothetic/dilatation and $Sp(1)$ Killing vectors are given by,
\begin{equation}
  \label{hkv}
  \chi^a = \tfrac{1}{2}\,  k_D^a = z^a\ , \quad
   k^{3 \, a} = \tfrac{1}{2} z^a\,,\ k^{+\, a} = 0\,,\
    k^{-\, a} = \tfrac{1}{2}\, (\Omega\, \bar z)^a\, .
\end{equation}
The conjugate components of the vectors are simply obtained by complex conjugation,
$k^{3\,\bar a} = (k^{3 \, a})^*$, $k^{+\, \bar a} = (k^{-\, a})^*$,  etc..
For the resulting triholomorphic isometries (\ref{3hol})
and moment maps (\ref{mmap}) one obtains,
\begin{align}
  \label{Xflat}
  X^a_{({\hat{\ww{m}}})} &\ = u_{({\hat{\ww{m}}})}{}^a{}_b\, z^b \quad \textrm{with} \quad
   u_{({\hat{\ww{m}}})}^\dagger = - u_{({\hat{\ww{m}}})} \ \  \textrm{and} \ \
   u_{({\hat{\ww{m}}})}^t = \Omega\, u_{({\hat{\ww{m}}})}\, \Omega \, , \nonumber\\[7pt]
   \mu^3_{({\hat{\ww{m}}})}&\ = i\, ( \bar z\,  u_{({\hat{\ww{m}}})}\, z)\ , \quad
    \mu^{(+)}_{({\hat{\ww{m}}})} = \tfrac{1}{2}\,(\Omega\, u_{({\hat{\ww{m}}})})_{ab}\, z^az^b\, .
\end{align}
The matrices $u_{({\hat{\ww{m}}})}$ thus generate the group $Sp(n)$, which is
the isometry group of the
underlying quaternionic manifold $\mathbb{H}P^{}$ (\ref{Qflat}). It is easy to see that
all isometries commute:
$[\, X_{\hat{\ww{m}}},\, k_D\,] =  [\, X_{\hat{\ww{m}}},\, \vec k\,] = [\, \vec k,\, k_D\,] =0$.
This would be not the case for the translational triholomorphic isometries of $\mathbb{R}^{4n}$.
However, the $Sp(1)$ R-isometry is obviously not manifestly
realized on the complex coordinates $z^a$, since it rotates the complex
structures (\ref{hkp}).

In order to realize the $Sp(1)$ R-isometry in a manifest way we
introduce the pseudo-real coordinates,
\begin{equation}
  \label{prc}
  q^{i\,a} = \begin{bmatrix} q^1\\ q^2\end{bmatrix} :=
  \kappa \begin{bmatrix} \Omega\, \bar z\\ -i\, z\end{bmatrix}
  =f^{ia}{}_\alpha\, q^\alpha \quad  \Rightarrow
  \quad (q^{i\,a})^* = \vep_{ij}\Omega_{ab}\, q^{j\,b}\, ,
\end{equation}
where $\kappa$ is a phase such that $\kappa^2=i$. The constant
flat vielbeine, $f^{ia}{}_\alpha$, $f^\alpha{}_{ia}$ are given explicitly in (\ref{flatVB}).
Dilatations, $Sp(1)$ and $Sp(n)$ action on these
coordinates are of the covariant form,
\begin{equation}
  \label{dq}
   \delta_{\lambda\, k_D} q^{i\,a} = 2\,\lambda\, q^{i\,a}\ ,\quad
 \delta_{\vec \alpha  \vec k}\, q^{i\,a} =
  \vec\alpha\,  \vec t^{\ i}{}_j\,  q^{i\,b}\ , \quad
 \delta_{\lambda^{\hat{\ww{m}}}\, X_{({\hat{\ww{m}}})}} q^{i\,a} =
  \lambda^{\hat{\ww{m}}}\, u_{({\hat{\ww{m}}})}{}^a{}_b\, q^{i\,b}\, ,
\end{equation}
where $\vec{t}^{\ i}{}_j =  -\tfrac{i}{2}\vec\sigma^{\,i}{}_j$. The coordinates $q^{i\,a}$ thus
transform in the fundamental representation $({\bf 2},{\bf 2n})$ under $Sp(1)\times Sp(n)$.
The basic data for the  HKC and the Lagrangian (\ref{La}) are given by
\begin{align}
  \label{psrctensors}
  g &\ =\ \vep_{ij}\,\Omega_{ab}\, dq^{i\,a}\, \otimes\,  dq^{j\,b}\ ,
  & \omega^{ij} &\ = \Omega_{ab}\, dq^{i\,a}\w dq^{j\,b}\ , \nonumber\\[7pt]
  X_{({\hat{\ww{m}}})}^{i\,a}&\ =\  u_{({\hat{\ww{m}}})}{}^a{}_b\, q^{i\,b}\ ,
  & \mu_{({\hat{\ww{m}}})}^{ij}&\ = (\,\Omega u_{({\hat{\ww{m}}})}\,)_{ab}\, q^{i\,a} q^{j\,b}\, .
\end{align}
The next step is to choose subgroups $G\subseteq Sp(n)$ and their representations
which can be embedded in the $sp(n)$ matrices $u_{({\hat{\ww{m}}})}$. Denoting the
corresponding matrices by $u_{(\ww{m})}$ one finds from (\ref{lie}), (\ref{prc}),
\begin{equation}
  \label{eq:4}
  [ \, u_{(\ww{m})}, u_{(\ww{n})}\, ] =  f_{\ww{m}\ww{n}}{}^{\ww{p}}\, u_{(\ww{p})}
 \;.
\end{equation}
\\
\noindent
{\bf{ADE embeddings.}}
We choose a canonical form for the $2n\times 2n$ component matrix $\Omega$ of the symplectic
form (\ref{vielbein}), (\ref{fHK}), which then specifies the general structure of any $sp(n)$
matrix $u$ (\ref{Xflat}):
\begin{equation}
  \label{spn}
  \Omega =\begin{bmatrix} & \unit\\ -\unit & \end{bmatrix} \quad \Rightarrow\quad
   u=\begin{bmatrix} A & B\\ -B^* & -A^t\end{bmatrix}\ \ \textrm{with:} \ \ A^\dagger =- A\, , \
   B^t = B\, .
\end{equation}
This allows for diverse embeddings of different groups. For example with $B=0$ one obtains the
embedding $U(n) \hookrightarrow Sp(n)$ with two copies of the hypermultiplets with the second
in the contragredient representation of the first one. A different
$U(n) = SO(2n)\cap Sp(n) \hookrightarrow Sp(n)$ embedding is obtained by by choosing
$A$ and $B$ to be real. If one sets $B=0$ in the latter case one obtains the
embedding $SO(n) \hookrightarrow Sp(n)$, again with two copies of hypermultiplets, one in
the contragredient representation of the other. It is thus clear that one can embed all
classical groups, including abelian factors, and taking the dimension $n$ large enough
any representation therof.

This means that at this stage there is no restriction to $ADE$ gauge groups as one would
expect for effective theories of multiple $M5$-branes. However, it is not to be expected
to happen at the classical level. In terms of effective CFT's the restriction to
$ADE$ gauge groups results from anomaly cancelation conditions \cite{Blum:1997mm},
and is thus an essential
quantum effect. We have to postpone such a discussion in the context of our models to
subsequent investigation.


\section{Superconformal Lagrangian}


\paragraph{Hypermultiplets.}

Supersymmetry requires the
tangent bundle of the hyper-K\"ahler cone $\ca{M}_{4n}$ to be of the form
$T\ca{M}_{(4n)} = \mathrm{H}_{Sp(1)}\otimes \mathrm{P}_{Sp(n)}$
\cite{Bagger:1983tt}, hence the structure group is  $Sp(1)\times Sp(n)$.
The sections of the trivial (pull-back) bundle $\mathrm{H}_{Sp(1)}$ are the constant
susy parameters $\epsilon^i$ and thus the $Sp(n)$ bundle $\mathrm{P}_{Sp(n)}$ defines the holonomy
group. This gives the mentioned  HK condition $\textrm{Hol}(g)\subseteq Sp(n)$.
Consequently there exists a local vielbein $f^{\alpha}{}_{ia}$ and its inverse
$f^{ia}{}_\alpha$ which satisfy the vielbein postulate and provide an expression for
the metric:
\begin{align}
  \label{vielbein}
  g_{\alpha\beta} = \vep_{ij}\,\Omega_{ab}\, f^{ia}{}_\alpha\, f^{jb}{}_\beta\quad  \Leftrightarrow&\ \quad
  g_{\alpha\beta}\, f^\alpha{}_{ia}\, f^\beta{}_{jb} = \vep_{ij}\, \Omega_{ab}\ ,\nonumber\\[7pt]
  \nabla_\alpha\, f^{ia}{}_\beta + \omega_\alpha{}^a{}_b\, f^{ib}{}_\beta = 0\quad\Leftrightarrow&\ \quad
  \delta^i{}_j\, \omega_\alpha{}^a{}_b = f^{ia}{}_\beta\nabla_\alpha\,f^\beta{}_{jb}\, .
\end{align}
Here, 
 $\omega_\alpha{}^a{}_b$ is the $Sp(n)$-connection.
We collect further properties of HK manifolds in appendix~\ref{Ageo}.

The susy transformations for the hypermultiplets are given by
\cite{Sierra:1983uh,Rosseel:2004fa},
\begin{align}
  \label{susy}
  \delta q^\alpha  =&\ f^{\alpha}{}_{ia}\,\bar\epsilon^i\,\psi^a \, ,\nonumber\\[5pt]
   \delta \psi^a =&\  \tfrac{1}{2}\, \slashed{D} q^\alpha\,\epsilon_i\,f^{ia}{}_\alpha
             - \delta q^\alpha\,\omega_\alpha{}^a{}_b\,\psi^b\, ,
\end{align}
where $D_\mu$ is the covariant derivative  (\ref{covder}).
The dilatation and special supersymmetry transformations
will be given below.
The conformally supersymmetric Lagrangian for the gauged sigma model coupled to the off-shell vector multiplet
$(\,A^{\ww{m}}_\mu,\,\lambda^{i\,\ww{m}},\, Y^{\ww{m}}_{ij}\,)$ can be obtained
by restricting the general supersymmetric model of~\cite{Sierra:1983uh}
to a HKC target space and the gauging of isometries satisfying (\ref{3hol}).
The resulting Lagrangian is \footnote{By a rescaling of the vector multiplet
$(\, A^{\ww{m}}, \lambda^{\ww{m}\, i}, Y_{ij}^{\ww{m}}\,)$
(and associated gauge parameters $\Lambda^{\ww{m}}$) one may introduce an
explicit coupling constant.}
\begin{align}
  \label{La}
  \ca{L_{\mathrm{hyp}}} = &\ - \tfrac{1}{2}\, g_{\alpha\beta}\, D_\mu q^\alpha D^\mu q^\beta
   + \bar\psi_a \gamma^\mu \mathscr{D}_\mu \psi^a
   - \tfrac{1}{8}\, W_{abcd}\, \bar\psi^a\gamma^\mu\psi^b\,\bar\psi^c\gamma_\mu\psi^d \nonumber\\[7pt]
    &\quad
    + 4\, \bar\psi_a\,{\bar\lambda}^{\ww{m}}_i\,f^{ia}{}_\alpha  X_{(\ww{m})}^\alpha
    +  Y^{\ww{m}}_{ij}\,\mu^{ij}_{(\ww{m})}\, ,
\end{align}
where $X^\alpha_{(\ww{m})}$ are the triholomorphic Killing vectors (\ref{lie}) and
$\mu_{(\ww{m})}{}^i{}_j = \imi\,\vec\sigma^{\,i}{}_j\,\vec\mu_{(\ww{m})}$ are the
associated moment maps
(\ref{mmap}). The $Sp(n)$ curvature tensor $W_{abcd}$ is defined in (\ref{geo5}). The
gauge covariant derivative for the fermions is given by
\begin{equation}
  \label{dtot}
  \mathscr{D}_\mu \psi^a \equiv
  \del_\mu\psi^a -  A^{\ww{m}}_\mu\, t_{(\ww{m})}{}^a{}_b\, \psi^b 
   + \del_\mu q^\alpha\,\omega_{\alpha}{}^a{}_b\,\psi^b
\;,
\end{equation}
where $t_{(\ww{m})}{}^a{}_b$ is defined as $t_{(\ww{m})}{}^a{}_b =\hal f^{ia}{}_{\alpha}\, \nabla_\beta X_{(\ww{m})}{}^\alpha\, f^\beta{}_{ib}$,
and satisfies the algebra~(\ref{eq:4}).
Under gauge transformations, (\ref{dtot}) transforms as
\begin{equation}
  \label{Dpsi}
  \delta_\Lambda\,\psi^a = L^a{}_b\, \psi^b\ ,
\quad
\delta_\Lambda (\, \mathscr{D}_\mu \psi^a\,)
   =  L^a{}_b\, \mathscr{D}_\mu \psi^b\ ,
     \quad
L^a{}_b \equiv   \left[\, \Lambda^{\ww{m}}\, t_{(\ww{m})}{}^a{}_b
     -\delta_\Lambda\, q^\alpha\,\omega_\alpha{}^a{}_b\,\right]
     \;.
\end{equation}
Finally, we recall the standard off-shell
supersymmetry transformations of the vector multiplet:
\be
\label{vm}
\delta A^\cm_\mu  = -\bar\epsilon\gamma_\mu\lambda^\cm\ ,
\quad \delta \lambda^{i\,\cm} = \tfrac{1}{8}\,\gamma^{\mu\nu} { F}^\cm_{\mu\nu} \epsilon^i
-\tfrac{1}{2}\,Y^{ij\,\cm}\epsilon_j\ ,
\quad
\delta Y^{ij\,\cm} \ = - {\bar\epsilon}^{(i}\slashed{D}\lambda^{j)\cm}\ .
\ee

\noindent
{\bf{Tensor-Vector multiplet.}} The Lagrangian (\ref{La}) is supersymmetric under
(\ref{susy}) and the standard off-shell susy transformation rules (\ref{vm}) 
for the vector multiplet.
The gauging should be accompanied by kinetic terms for the vector multiplet,
however the pure Yang-Mills action is not conformally invariant in six dimensions.
To achieve conformal invariance a compensating supermultiplet is needed. This role
can be played by a collection of tensor multiplets and the corresponding model
has been constructed in~\cite{Samtleben:2011fj}. The field content of this conformal
tensor-vector model is given by a set of $n_{\rm T}$ tensor multiplets
$\{\phi^I, B_{\mu\nu}^I, \chi^I\}$ and $n_{\rm V}$ vector multiplets
$\{A_\mu^r, \lambda^r, Y_{ij}^r\}$, labeled by indices $I$ and $r$, respectively.
In addition, the model in its most general form 
requires the introduction of the non-propagating 3-form potentials $C_{\mu\nu\rho\, r}$.
The vectors $A^{\ww{m}}$ gauging the HKC isometries, cf.\ (\ref{covder}), and its superpartners
will be identified with a subset of these fields below. To begin with, we shall recall the
superconformal invariant interactions of these multiplets that admit an action formulation
(modulo the standard subtleties of actions for self-dual tensor fields), 
as constructed in \cite{Samtleben:2011fj} to which we refer for details.
We shall then discuss the superconformal invariance of  the total action which also includes
the hypermultiplets.

For vector and two-form tensor fields, the full covariant non-abelian field strengths are given by
\begin{eqnarray}
{\cal F}_{\mu\nu}^r &\equiv&
2 \partial_{[\mu} A_{\nu]}^r - f_{st}{}^r A_\mu^s A_\nu^t + h^r{}_I\,B_{\mu\nu}^I
\;,\nonumber\\[.5ex]
{\cal H}_{\mu\nu\rho}^I &\equiv& 3 D_{[\mu} B_{\nu\rho]}^I +
6 \, d^I{}_{rs}\,  A_{[\mu}^r \partial^{\vphantom{r}}_\nu A_{\rho]}^s
- 2 f_{pq}{}^s d^I{}_{rs}\, A_{[\mu}^r A_\nu^p A_{\rho]}^q
+ h^{rI} C_{\mu\nu\rho\,r}\ ,
\label{defF}
\end{eqnarray}
in terms of the antisymmetric structure constants $f_{st}{}^r=f_{[st]}{}^r$,
a symmetric $d$-symbol $d^I{}_{rs}=d^I{}_{(rs)}$, and the tensor $h^r{}_I$
inducing general St\"uckelberg-type couplings among forms of different degree.
Indices $I, J$ are raised and lowered with a constant symmetric tensor $\eta_{IJ}$.
The covariant derivatives are defined as $D_\mu \equiv \partial_\mu - A_\mu^r X_r$.
The field strengths (\ref{defF}) are defined such that they transform covariantly under the
set of non-abelian gauge transformations
\begin{eqnarray}
\delta A_\mu^r &=& D_\mu \Lambda^r - h^r{}_I \Lambda_\mu^I
\;,\nonumber\\[.5ex]
\Delta B_{\mu\nu}^I &=& 2 D_{[\mu} \Lambda_{\nu]}^I -2\, d^I{}_{rs} \,\Lambda^r {\cal F}_{\mu\nu}^s
- h^{rI} \Lambda_{\mu\nu\,r}
\;,\nonumber\\[.5ex]
h^r{}_I \Delta C_{\mu\nu\rho\,r} &=& h^r{}_I\left( 3 D_{[\mu} \Lambda_{\nu\rho]\,r}
+6 \, d_{Irs}\,{\cal F}_{[\mu\nu}^s \,\Lambda_{\rho]}^I
+ 2d_{Irs}\,{\cal H}_{\mu\nu\rho}^I \,\Lambda^s\right) \ ,
\label{gaugesym}
\end{eqnarray}
where we have used the compact notation
\begin{eqnarray}
\Delta B^I_{\mu\nu} &\equiv& \delta B^I_{\mu\nu} - 2 d^I{}_{rs}\,A_{[\mu}^r \,\delta A_{\nu]}^s
\;,\nonumber\\
\Delta C_{\mu\nu\rho\,r} &\equiv& \delta  C_{\mu\nu\rho\,r}
-6\, d_{Irs}\,B_{[\mu\nu}^I \,\delta A_{\rho]}^s
-4\, d_{Irs}\, d^I{}_{pq}\,A_{[\mu}^s \,A_{\nu}^p \,\delta A_{\rho]}^q\ .
\label{Delta1}
\end{eqnarray}
This tensor/vector gauge system is completely defined by the choice of the
constant tensors $h^r{}_I$, $d^I{}_{rs}$, $f_{rs}{}^t$.  Consistency of the tensor hierarchy, i.e.\ covariance of the field strengths (\ref{defF}) requires that the gauge group generators in the various representations are given by
\begin{eqnarray}
(X_{r})_{s}{}^t &=&(X^{\rm V}_{r})_{s}{}^t~\equiv~ -f_{rs}{}^t + h^t{}_I\,d^I{}_{rs}\ ,
\nonumber\\
(X_{r})_{I}{}^J &=& (X^{\rm T}_{r})_{I}{}^J~\equiv~ 2\, d^J{}_{rs} h^s{}_I -2h^{sJ} d_{Isr}\ ,
\label{genpar}
\end{eqnarray}
in terms of the constant tensors parametrizing the system. Further constraints 
on these tensors follow from closure of the algebra (or generalized Jacobi identities)
\begin{eqnarray}
[X_r, X_s] &=& -(X_{r})_{s}{}^t \,X_t\ ,
\label{jacobi}
\end{eqnarray}
and gauge invariance of the tensor $d^I{}_{rs}$,
see~\cite{Samtleben:2011fj} for the explicit form of these constraints.
The `action' describing the superconformal invariant couplings of the non-abelian tensor/vector 
system is given by 
\begin{align}
  \label{SUS6}
  {\cal L}_{\mathrm{VT}} &=
\tfrac{1}{8} d_{I rs}\, \phi^I
\left(
{\cal F}_{\mu\nu}^r {\cal F}^{\mu\nu\, s}
-4\,Y_{ij}^{r} Y^{ij\,s} + 8 \bar\lambda^r \slashed{D} \lambda^s
\right)
-\tfrac{1}{8} D^\mu \phi_I \,D_\mu \phi^I
-\tfrac{1}{2} \bar\chi_I\, \slashed{D} \chi^{I}
\nonumber\\[7pt]
&{}
-\tfrac{1}{96}  {\cal H}_{\mu\nu\rho}^I\, {\cal H}^{\mu\nu\rho}_I
-\tfrac{1}{24} d_{Irs} {\cal H}_{\mu\nu\rho}^I\,\bar\lambda^r\gamma^{\mu\nu\rho}  \lambda^{s}
- \tfrac12 d_{Irs}\left(\, {\cal F}_{\mu\nu}^r\,\bar\lambda^s\gamma^{\mu\nu}  \chi^{I}
-4 Y_{ij}^r\,\bar\lambda^{i\,s}\chi^{j\,I} \,\right)
\nonumber\\[7pt]
&{}
+  \left( d_{J sr} h^{s}{}_I  -4 d_{I sr} h^{s}{}_J \right) \phi^I  \bar\lambda^r \chi^J
+  \tfrac{1}{4}  d_{Irs} h^r{}_Jh^s{}_K\,\phi^I \phi^J\phi^K
\nonumber\\[7pt]
&{}
-\tfrac{1}{6}
d_{Irs} d^I_{uv}\,\bar\lambda^r\gamma^\mu \lambda^{u} \bar\lambda^s \gamma_\mu \lambda^v
-\tfrac{1}{48} {\cal L}_{\rm top}\;,
\end{align}
where the topological term ${\cal L}_{\rm top}$ is given by integrating
\bea
dV\delta {\cal L}_{\rm top}
= 6\left\{
2d_{I r s}\,\delta A^r\w {\cal F}^s\w {\cal H}^{I}
 -\Delta B^I\w\left( h^r{}_I\,{\cal H}^{(4)}_r
- d_{Irs} {\cal F}^r\w{\cal F}^s\right)
- h^r{}_I \Delta C_{r}\w {\cal H}^{I}\right\}\ ,
\nonumber
\eea
with ${\cal H}^{(4)}$ defined in~(\ref{TH5}) below. 
Finally, it is important to note that for the tensor multiplet, this action has to be supplemented with the first-order self-duality equation
\begin{eqnarray}
{\cal H}^{I\,-}_{\mu\nu\rho} &=&-  d^I_{rs} \bar\lambda^r \gamma_{\mu\nu\rho} \lambda^s\ .
\label{firsto}
\end{eqnarray}
to be imposed {\em after} having derived the second-order equations of motion. This is due to the well known fact that $p$-form potentials with self-dual field strengths do not admit a manifestly Lorentz covariant action formulation.
Moreover, it has been observed in~\cite{Samtleben:2011fj} that the metric $\eta_{IJ}$ defining the kinetic terms in the tensor sector
is necessarily of indefinite signature, i.e.\ a priori the spectrum of the Lagrangian~\ref{SUS6} contains ghosts
whose fate will require a more extensive analysis.

The action of the above Lagrangian is invariant under the following supersymmetry transformations 
\begin{align}
\label{vmmod}
\delta A^r_\mu&\ = -\bar\epsilon\gamma_\mu\lambda^r\;,
\nonumber\\[.5ex]
\delta \lambda^{i\,r} &\ = \tfrac{1}{8}\,\gamma^{\mu\nu} {\cal F}^r_{\mu\nu} \epsilon^i
-\tfrac{1}{2}\,Y^{ij\,r}\epsilon_j + \tfrac{1}{4}\,h^r{}_I \phi^I \epsilon^i \;,
\nonumber\\[.5ex]
\delta Y^{ij\,r} &\ = - {\bar\epsilon}^{(i}\slashed{D}\lambda^{j)r}
+2\,h^r{}_I\,\bar\epsilon^{(i} \chi^{j) I}\, ,
\nonumber\\[1ex]
\delta \phi^I\, &\ = \bar\epsilon\chi^I\;,
\nonumber\\[.5ex]
\delta\chi^{i\,I}\, &\ = \tfrac{1}{48} \, \gamma^{\mu\nu\rho} \,{\cal H}^{I}_{\mu\nu\rho}
\epsilon^i +\tfrac{1}{4}\,\slashed{D}\phi^I \epsilon^i
+ \tfrac{1}{2} d^I{}_{rs} \bar\lambda^{i\,r}\gamma^\mu\lambda^{j\, s}\,\gamma_\mu\epsilon_j \;,
\nonumber\\[.5ex]
\Delta B^I_{\mu\nu}\, &\ = -\bar\epsilon\gamma_{\mu\nu}\chi^I\;,
\nonumber\\[.5ex]
h^r{}_I \Delta C_{\mu\nu\rho\,r}\, &\ = -2h^r{}_I d_{Jrs} \left(\bar\epsilon\gamma_{\mu\nu\rho}\lambda^s \phi^J\right)\ .
\end{align}
Next, we turn to the description of the full action describing the superconformal couplings of the hypermultiplets to the non-abelian tensor/vector system.
\\

\noindent
{\bf{The full Lagrangian.}} To put together the actions for the hypermultiplet and the tensor/vector system, we need to identify the vectors $A_\mu^\cm$ used to gauge the HKC isometries, cf. (2.10), as well as its superpartners, as a subset of the the vectors $A_\mu^r$ and their superpartners according to
\begin{equation}
\label{embt}
A^{\ww{m}} =  A^r\theta_r{}^{\ww{m}}\ ,\qquad \lambda^\cm =\lambda^r \theta_r{}^\cm\ ,
\qquad Y_{ij}^\cm = Y_{ij}^r\theta_r{}^\cm\ ,
\end{equation}
with the constant embedding tensor $\theta_r{}^\cm$.
For consistency, the embedding of the vector fields~(\ref{embt})
has to be supplemented with the following constraints on the
embedding tensor:
\begin{equation}
 h^r{}_I\,\theta_r{}^{\ww{m}} ~{=}~  0\;, \qquad f_{rs}{}^t\, \theta_t{}^{\ww{m}}
~{=}~
\theta_r{}^{\ww{n}}\, \theta_s{}^{\ww{p}} \, f_{\ww{n}\ww{p}}{}^{\ww{m}}\ .
\label{ec}
\end{equation}
The first condition guarantees that the modification of the
susy transformations (\ref{vmmod}) does not affect the variation of the hypermultiplet action.
Whereas the second condition guarantees that the embedding is homomorphic.
The full Lagrangian is given by the sum
\bea
{\cal L} &=& {\cal L}_{\rm VT} + \frac1{2\lambda} {\cal L}_{\rm hyp}\ .
\label{Lagtotal}
\eea
where we introduced a relative (dimensionless) coupling constant $\lambda$.
Both Lagrangians are separately supersymmetric so that $\lambda$ is a free parameter.

The fact that both actions are separately supersymmetric requires the first of the conditions in (\ref{ec}).
Note also that while the field equations of the tensor multiplets do not involve the hypermultiplet, those of the vector- and hypermultiplets evidently mix. The invariance of the total action guaranties the fact that all the field equation transform into each other under supersymmetry. Nonetheless, given the fact that the field equation for the auxiliary field $Y_r^{ij}$ contains a contribution coming from the moment map of the hypermultiplet sector, it is instructive to examine how the vector multiplet field equations behave under supersymmetry. These field equations take the form
\bea
\delta Y_r^{ij}: \qquad && {\cal E}_r^{ij} + \frac1{2\lambda}\,\theta_r{}^\cm \mu_{(\cm)}^{ij}=0\ ,
\label{ym1}\\[.5ex]
\delta \lambda^{ir}:\qquad && {\cal E}_r^i + \frac2{\lambda} \,\theta_r{}^\cm X_{(\cm)}^\alpha f_\alpha^{ia} \psi_a=0\ ,
\label{ym2}\\[.5ex]
\delta A_\mu^r: \qquad && {\cal E}_{\mu r} + \frac1{2\lambda}\,\theta_r{}^\cm \left( D_\mu q^\alpha X_{(\cm)\alpha} -t_{(\cm)}{}^a{}_b {\bar\psi}_a\gamma_\mu\psi^b\right)=0\ ,
\label{ym3}
\eea
where ${\cal E}_r^{ij}, {\cal E}_r^i$ and ${\cal E}_{\mu r}$ represent the contributions from the Lagrangian ${\cal L}_{\rm VT}$ to the field equations of the vector multiplet. These contributions are established to transform into each other in \cite{Samtleben:2011fj}, and one can check that the supersymmetry variation of the contributions multiplying the embedding tensor $\theta_r{}^\cm$ also transform into each other, as expected.


\paragraph{Higher $p$-forms and duality equations.}


It has been shown in~\cite{Samtleben:2011fj} that the tensor/vector system described by (\ref{SUS6})
can be extended on-shell to include higher order $p$-forms that are related by first-order duality
equations to the physical fields. Let us briefly discuss how this extension is modified in the presence of
hypermultiplets.

The $p$-form field content of (\ref{SUS6}) is given by vector and tensor fields $A_\mu{}^r$ and 
$B_{\mu\nu}{}^I$ and the three-form potentials $C_{\mu\nu\rho\,r}$ which however only appear 
under projection with the tensor $h^r{}_I$\,. In a first step, the model may be extended to 
the full set of three-forms $C_{\mu\nu\rho\,r}$,
as well as four-form potentials $C^{(4)}_{\mu\nu\rho\sigma\,\ww{m}}$.
Supersymmetry and gauge transformations of the former are given by
\bea
\Delta C_{\mu\nu\rho\,r}\, &=& -2
d_{Jrs} \left(\bar\epsilon\gamma_{\mu\nu\rho}\lambda^s \phi^J\right)\ ,
\nonumber\\
\Delta C_{\mu\nu\rho\,r} &=& 3 D_{[\mu} \Lambda_{\nu\rho]\,r}
+6 \, d_{Irs}\,{\cal F}_{[\mu\nu}^s \,\Lambda_{\rho]}^I
+ 2d_{Irs}\,{\cal H}_{\mu\nu\rho}^I \,\Lambda^s
-\theta_r{}^{\ww{m}}\,\Lambda_{\mu\nu\rho\,{\ww{m}}}
\;,
\label{DeltaC}
\eea
respectively. Here, $\theta_r{}^{\ww{m}}$ is the embedding tensor of (\ref{embt})
and $\Lambda_{\mu\nu\rho\,{\ww{m}}}$ is the gauge parameter of the four-form
potentials. Closing the supersymmetry algebra on the three-form potentials leads 
to the field equations (\ref{ym1}). The hypermultiplet contribution of (\ref{ym1}) is now absorbed into 
the resulting St\"uckelberg transformation $\Lambda_{\mu\nu\rho\,{\ww{m}}}$ on the three-form
potential which explains the appearance of the embedding tensor $\theta_r{}^\cm$
in (\ref{DeltaC}).
Furthermore, closure of the supersymmetry algebra implies
the first-order duality equation
\bea
  2d_{Irs}\left(\,{\cal F}_{\mu\nu}^s \phi^I-2\,\bar\lambda^s \gamma_{\mu\nu} \chi^I\, \right)
&=& \tfrac{1}{4!}\,\varepsilon_{\mu\nu\lambda\rho\sigma\tau}\,{\cal H}^{(4)\,\lambda\rho\sigma\tau}_{r}
\;.
\label{FH}
\eea
where the field strength ${\cal H}^{(4)}_r$ is defined as
\begin{align}
  \label{TH5}
  \ca H^{(4)}_{r} &= DC_r + \theta_r{}^{\ww{m}}\, C^{(4)}_{\ww{m}}
  - 2d_{Irs}\left (\, \ca F^s\w B^I-\tfrac{1}{2}\, h^s{}_J\, B^I\w B^J\right.
   \nonumber\\[.5ex]
  &\hspace{4cm} \left. +\tfrac{1}{3} d^I{}_{pq}\, A^s\w K^{rq} 
    + \tfrac{1}{36}\, f_{uv}{}^q\, d^I{}_{pq}\, A^s\w A^p\w A^u\w A^v\,\right)\;.
\end{align}
In turn, it satisfies the Bianchi identity
\begin{equation}
  \label{TH7}
  D\ca H^{(4)}_r = -  2\,d_{Irs}\,\ca F^s\,\w \ca H^I +  \theta_r{}^{\ww{m}}\,\ca H^{(5)}_{\ww{m}}\, ,
\end{equation}
with ${\cal H}^{(5)}_{\ww{m}}$ denoting the field strength of the 
four-form potentials $C^{(4)}_{\ww{m}}$.
Equation~(\ref{FH}) is the non-abelian first-order duality equation that relates the
three-form potentials $C_{\mu\nu\rho\,r}$ to the vector fields and unlike the field equations
(\ref{ym1})--(\ref{ym3}) has no contributions from the hypermultiplet.
Nevertheless, its derivative is precisely compatible with the second-order Yang-Mills equation (\ref{ym3}),
provided the field strength ${\cal H}^{(5)}_{\ww{m}}$ in turn satisfies the first-order duality equation
\bea
\tfrac{1}{5!}\,\varepsilon_{\mu\nu\rho\lambda\sigma\tau}\,
\theta_r{}^\cm\,{\cal H}^{(5)\,\nu\rho\lambda\sigma\tau}_{\ww{m}}
&=&
   \left[\, (X_{r})_{IJ}
  \left(\phi^I D_\mu \phi^J-2\bar\chi^I \gamma_\mu \chi^J \right)
+4(X_r)_u{}^s\, d_{Isv}\,\phi^I\,\bar\lambda^u\gamma_\mu \lambda^v\, \right]
\nonumber\\
&&{}+
\frac{2}{\lambda}
    \theta_r{}^\cm \left(  X_{(\ww{m})\,\alpha}\, D_\mu q^\alpha 
     - \bar\psi_a\,\gamma_\mu\,t_{(\ww{m})}{}^a{}_b\,\psi^b\right)\; ,
\eea
relating the four-form potentials $C^{(4)}_{\ww{m}}$ 
to the scalar fields of the model (including hypers).
Accordingly, the supersymmetry transformation rule for the 4-form potentials 
in presence of hypermultiplets is modified to
\be
\theta_r{}^\cm \Delta C_{\mu\nu\rho\sigma\, \cm} = 
(X_{r})_{IJ}\,\phi^{[I}\,{\bar\epsilon}\gamma_{\mu\nu\rho\sigma}\chi^{J]} 
- \frac{2}{\lambda} \,\theta_r{}^\cm\, X_{(\cm)}^\alpha f^{ia}{}_\alpha 
\,{\bar\epsilon}_i\gamma_{\mu\nu\rho\sigma} \psi_a \;,
\label{susyC4}
\ee
combining contributions from tensor and hypermultiplets.


\paragraph{Superconformal symmetries.}


We conclude with a presentation of the superconformal symmetry transformations \cite{Sezgin:1994th}. Denoting the fields in the theory by $\Phi=(\phi^I, B_{\mu\nu}^I,\chi^I,A_\mu^r, Y^{ij},\lambda^r, C_{\mu\nu\rho \,r})$, the conformal transformations are given by
\be
\delta_C \Phi = {\cal L}_\xi \Phi  + \lambda_D \Omega \Phi\ ,
\ee
where ${\cal L}_\xi$ is the Lie derivative with respect to the conformal Killing vector defined by
$\partial_{(\mu} \xi_{\nu)}=\Omega\eta_{\mu\nu}$, which also defines $\Omega$,
and $\lambda_D$ are the Weyl weight for $\Phi$ given by $(2,0,5/2,0,2,3/2,0)$. The Lie derivative for the fermionic fields $\Psi=(\psi^a, \chi^I, \lambda^r)$, in particular, takes the form ${\cal L}_\xi \Psi = \xi^\mu\partial_\mu\Psi +\frac14 \partial_\mu\xi_\nu \gamma^{\mu\nu}\Psi$. On the other hand, the hypermultiplet fields transform as
\bea
\delta_C q^\alpha &=& {\cal L}_\xi q^\alpha +  4 \Omega \chi^\alpha\ ,
\nonumber\\
\delta_C \psi^a &=& {\cal L}_\xi  \psi^a +\frac52 \Omega \psi^a - \Omega \chi^\alpha \omega_\alpha{}^{a}{}_b\,\psi^b\ ,
\eea
with the homothetic Killing vector and spin connection from (\ref{hKV}) and (\ref{vielbein}),
respectively.
The conformal supersymmetry transformations, on the other hand, involve conformal Killing spinors, consisting of a pair $(\eta_+, \eta_-)$ that satisfy $\partial_\mu\eta_+ -\frac12 \gamma_\mu\eta_-=0$. The superconformal transformations take the form of supersymmetry transformations in which the constant supersymmetry parameter $\epsilon$ is replaced by $\eta_+$, and the parameter $\eta_-$ arises only in $\delta_{\eta_-} \chi^I = -\frac12 \phi^I \eta_-$.
Note that the bosonic conformal transformation together with supersymmetry
ensures the full superconformal symmetry since the commutator of conformal boost with
supersymmetry yields the special supersymmetry generator \cite{Bergshoeff:1985mz}.


\section{Eliminating auxiliary fields}


In the previous sections we have coupled the hypermultiplet Lagrangian of~\cite{Sierra:1983uh}
to the tensor/vector system of~\cite{Samtleben:2011fj}.
The non-trivial interactions between vector-, tensor- and hypermultiplets are reflected
in equation (\ref{ym1})
\bea
d_{I\,rs}\,\phi^I \, Y^{ij\,s} - \frac1{2\lambda}\,\theta_r{}^\cm \mu_{(\cm)}^{ij}&=& 2\, d_{I\,rs} \, \bar\lambda^{s(i} \chi^{j)I}
\;,
\label{aux}
\eea
for the auxiliary fields $Y^{ij\,s}$ of the Yang-Mills multiplet.
In particular, eliminating the auxiliary fields will introduce non-trivial couplings between tensor- and hypermultiplets
of the model.

In this section, we will further analyze the form and the consequences of equation~(\ref{aux})
and derive the couplings induced by elimination of the auxiliary fields.
The explicit form of ~(\ref{aux}) depends on the field content of the model and 
the particular choice of the constant 
tensors $d_{I\,rs}$, $\theta_r{}^\cm$, parametrizing the model. As discussed above, these constant 
tensors are subject to a number of algebraic constraints derived from (\ref{jacobi}).
A general class of solutions to these constraints has been constructed in~\cite{Samtleben:2012mi} 
based on a semi-simple Lie algebra $\mathfrak{g}$ under
which all fields transform in non-trivial representations. Explicitly, w.r.t.\ this algebra the vector and tensor multiplets split into
\bea
&& A_\mu^r \to \left( A_\mu^\cm, B_\mu^{\rm A}\right)\ ,
\qquad\ \  \lambda^r \to \left(\lambda^\cm, \nu^{\rm A}\right)\ ,
\qquad Y^r_{ij}\ \  \to \left(Y^\cm_{ij}, Z^{\rm A}_{ij}\right)\ ,
\nonumber\\[0.5ex]
&& B_{\mu\nu}^I \to \left(B_{\mu\nu}^{\rm A}, C_{\mu\nu\,{\rm A}}\right)\ ,
\qquad \chi^I \to \left(\chi^{\rm A}, \zeta_{\rm A}\right)\ ,
\qquad \phi^I \to \left(\phi^{\rm A}, \varphi_{\rm A}\right)\ ,
 \label{split}
\eea
where indices ${\ww{m}}$ and ${\rm A}$ refer to the adjoint and an arbitrary fixed 
representation ${\cal R}$ of~$\mathfrak{g}$, respectively. 
The model thus combines $n_{\rm V}={\rm dim}\,\mathfrak{g} + {\rm dim}\,{\cal R}$ vector multiplets
and $n_{\rm T}=2\, {\rm dim}\,{\cal R}$ tensor multiplets.
The non-vanishing components of the gauge invariant constant tensors that define the model
(referred to as Type III in \cite{Samtleben:2012mi}) are
\bea
\eta_{\rm A}{}^{\rm B} &=&\eta^{\rm B}{}_{\rm A} ~=~\delta_{\rm A}^{\rm B}\ ,\quad
h^{\rm B}{}_{\rm A}~=~ \delta_{\rm A}^{\rm B}\ ,
\nonumber\\[.5ex]
f_{\cm {\rm A}}{}^{\rm B} &=& -\frac12 (T_\cm)_{\rm A}{}^{\rm B}\ ,\qquad f_{\cm\ww{n}}{}^{\ww{p}}\ ,
\nn\\[.5ex]
d^{\rm B}{}_{\ww{m} {\rm A}}&=&\frac12 (T_\ww{m})_{\rm A}{}^{\rm B}\ ,\quad
d_{\rm ABC} = d_{({\rm ABC})}\ ,\quad d_{{\rm AB}\ww{m}}= d_{({\rm AB})\ww{m}}\ ,
\quad d_{{\rm A} \ww{m}\ww{n}}\ .
\label{dtensors}
\eea
Here, $f_{\cm\ww{n}}{}^{\ww{p}}$, and $(T_\cm)_{\rm A}{}^{\rm B}$ denote the structure constants and
representation matrices of $\mathfrak{g}$, respectively, and
$d_{({\rm ABC})}$, $d_{({\rm AB})\ww{m}}$, $d_{{\rm A} \ww{m}\ww{n}}$,
are $\mathfrak{g}$-invariant tensors with the indicated symmetry properties,
which obviously exist only for particular choice of $\mathfrak{g}$ and ${\cal R}$.
In particular, the cubic scalar potential of~(\ref{SUS6})  
is exclusively triggered by the constant tensor $d_{\rm ABC}$\,:
\bea
{\cal L}_{\rm pot} &\propto& d_{\rm ABC} \,\phi^{\rm A} \phi^{\rm B}\phi^{\rm C}
\;.
\label{potential}
\eea
For the hypermultiplet couplings, we choose the
embedding tensor $\theta_r{}^ \cm$ as
\bea
\theta_\ww{m}{}^\ww{n}=\delta_\ww{m}{}^\ww{n}\;,\qquad
\theta_{\rm A}{}^\ww{n}=0\;,
\eea
such that only the vector fields $A_\mu^\ww{m}$ participate in the gauging,
and the algebra $\mathfrak{g}$ is identified with the algebra of gauged isometries
(\ref{lie}) in the hyper-sector.
In contrast, the vector fields $B_\mu^{\rm A}$ as well as the tensors $C_{\mu\nu\,{\rm A}}$
can be eliminated from the Lagrangian by field redefinition (see~\cite{Samtleben:2012mi} for details).

As we have discussed in the introduction, a particularly interesting class of models
is supposed to describe a field content that can be regrouped into multiplets of 
$(2,0)$ supersymmetry. In particular, this requires that tensor- and hypermultiplets arise
in the same representation (i.e.\ $n_{\rm H}=n_{\rm T}$), 
such that they may recombine into $(2,0)$ tensor multiplets.
The algebra $\mathfrak{g}$ is embedded into the orthogonal group $SO(n_{\rm T})$
via the generators $(X_\cm)_{IJ}$ from (\ref{genpar}).
Correspondingly, in this case its action in the hypermultiplet sector is realized via the embedding 
(\ref{spn}) (with $B=0$) into the $SO(n_{\rm T})$ subalgebra of isometries on the flat target space of hypermultiplets.
Indeed, with this realization it follows that part of the structures such as the supersymmetry
variation (\ref{susyC4}) combining tensor- and hyper-scalars can be embedded into the manifest $(2,0)$ form
\bea
 \Delta C_{\mu\nu\rho\sigma\, \cm} &=& 
(X_{\cm})_{IJ}\,\phi^{\hat\imath\hat\jmath,I}\,{\bar\epsilon}_{\hat\imath}\gamma_{\mu\nu\rho\sigma}\chi_{\hat\jmath}^{J} 
\;,
\label{susy_enh}
\eea
with $Sp(2)$ R-symmetry indices $\hat\imath, \hat\jmath = 1, \dots, 4$, and the 
five scalar fields $\phi^{[\hat\imath\hat\jmath]}$
combining the tensor- and hyper-scalars.
Truncating the (2,0) susy parameter as 
$\epsilon^{\hat\imath}\rightarrow(\epsilon^i, \tilde\epsilon^i) \rightarrow (\epsilon^i, 0)$, 
equation (\ref{susy_enh}) indeed reduces to (\ref{susyC4}).

In the context of the M5-brane dynamics, the most interesting models describe tensor multiplets
in the adjoint representation of the gauge group, i.e.\ correspond to the choice ${\cal R}={\rm Adj}_{\mathfrak{g}}$\,.
For the rest of this section, we will study a slightly more general class corresponding to choosing 
${\cal R}={\rm Adj}_{\mathfrak{g}} \oplus {\bf 1}$.
The role of the extra singlet will become clear in the following.
Explicitly thus, tensor multiplets split into
\bea
(B_{\mu\nu}^\cm, \chi^\cm, \phi^\cm)\;,
(B_{\mu\nu}^0, \chi^0, \phi^0)\;,
(C_{\mu\nu \cm}, \zeta_\cm, \varphi_\cm)\;,
(C_{\mu\nu 0}, \zeta_0, \varphi_0)\;.
\eea
Furthermore, for the constants in (\ref{dtensors}),
we choose the non-vanishing $\mathfrak{g}$-invariant tensors
\bea
d_{{\rm A}\ww{mn}} &:&
\left\{
\begin{tabular}{rl}
$c_1\,\delta_{\ww{m}\ww{n}}$ & for  ${\rm A}=0\;,$ 
\\
$0$ & otherwise \;,
\end{tabular}
\right.
\nonumber\\
d_{{\rm AB}\ww{m}} &:&
\left\{
\begin{tabular}{rl}
$c_2\,\delta_{\ww{m}\ww{n}}$ & for  $({\rm A}, {\rm B})=(0,\ww{n})$ or $({\rm A}, {\rm B})=(\ww{n},0)\;,$
\\
$0$ & otherwise \;,
\end{tabular}
\right.
\label{mind}
\eea
and set to zero all components of $d_{({\rm ABC})}$\,.
In particular, absence of $d_{({\rm ABC})}$ implies that there is no (unstable) cubic potential (\ref{potential})
for the tensor scalars.
The part of the total Lagrangian containing the auxiliary fields then takes the form (suppressing R-symmetry indices)
\bea
{\cal L}_Y &=& {\rm Tr}\Big[\,  Y[Z,\varphi]  -c_1\phi^0 Y^2 -2c_2 \phi^0 YZ +\frac{2}{\lambda} \mu Y\
\nonumber\\[.07ex]
&& -2Y\{ {\bar\nu},\zeta\} +2 Z\{ {\bar\lambda},\zeta\}  +4c_1 Y{\bar\lambda}\chi^0 +4c_2 Y {\bar\nu}\chi^0 +4c_2 Z{\bar\lambda}\chi^0\,\Big]\ ,
\label{Laux}
\eea
where all the fields are matrix valued and in the adjoint representation of $\mathfrak{g}$\,. 
The resulting field equations (\ref{aux}) for the auxiliary fields can be written as
\bea
{} [Y,\varphi]+2c_2\phi^0 Y &=& J\ ,
\label{yeq1}\\
{} [Z,\varphi] -2\phi^0 (c_1Y + c_2Z) +\frac{2}{\lambda} \mu &=& K\ ,
\label{yeq2}
\eea
where $J$ and $K$ are bilinear in fermions which can be easily read off from (\ref{Laux}).
The form of these equations shows that for generic values of the parameters $c_{1,2}$, the auxiliary fields
$Y$ and $Z$ are uniquely determined and can be eliminated from the Lagrangian. 
On the other hand, for $c_{1,2}=0$
only part of $Y$ and $Z$ is determined which implies constraints on the sources $J$ and $K$.
To make this explicit, it is convenient to consider the Lie algebra commutators 
in the Cartan-Weyl basis, in which the generators are denoted by $({\vec H}, E^\alpha,E^{-\alpha})$. 
Furthermore let us take the field $\varphi$ to lie in the Cartan subalgebra as
\be
\varphi = \vec \varphi \cdot \vec H \ .
\label{csa}
\ee
The non-vanishing commutators are
\be
[\vec H , E^{\pm\alpha}]= \pm \vec\alpha\, E^{\pm\alpha}\ ,\quad
[E^\alpha\ , E^\beta] = N_{\alpha\beta} E^{\alpha+\beta}\ , \quad
[E^\alpha\ , E^{-\alpha}] = \frac{2}{|\alpha|^2}\,\vec \alpha \cdot \vec H\ ,
\ee
where $\vec\alpha$ is the root vector and $N_{\alpha\beta}$ are numbers associated with the specific Lie algebra. Thus, expanding
\be
Y= \vec Y \cdot \vec H + Y_\alpha E^\alpha + Y_{-\alpha} E^{-\alpha}\ ,\qquad {\rm idem}\ Z, J, K\ ,
\label{exp}
\ee
from (\ref{yeq1}) it readily follows that~\footnote{The fields $(Y,Z,J,K, \mu)$ are understood to be rotated by the similarity transformation that puts the scalar field $\varphi$ in the Cartan subalgebra in the Cartan-Weyl basis.}
\be
Y = \frac{1}{2c_2\phi^0}\,\vec J \cdot \vec H + \frac{J_\alpha E^\alpha}{2c_2\phi^0+\vec\alpha\cdot\vec\varphi}
+ \frac{J_{-\alpha} E^{-\alpha}}{2c_2\phi^0-\vec\alpha\cdot\vec\varphi}\ .
\label{s1}
\ee
Solving (\ref{yeq2}) similarly and substituting the solution for $Y$ we then find
\bea
Z &=& -\frac{1}{2c_2\phi^0} \left(\,\vec K + 2c_1\phi^0 \vec J -\frac{2}{\lambda}\vec \mu\,\right)\cdot \vec H
- \frac{1}{2c_2\phi^0-\vec\alpha\cdot\vec\varphi}
\left(K_\alpha +\frac{2c_1\phi^0 J_\alpha}{2c_2\phi^0+\vec\alpha\cdot\vec\varphi} -\frac{2}{\lambda}\mu_\alpha \,\right)\,E^\alpha
\nonumber\\
&&- \frac{1}{2c_2\phi^0+\vec\alpha\cdot\vec\varphi}
\left(K_{-\alpha} +\frac{2c_1\phi^0 J_{-\alpha}}{2c_2\phi^0-\vec\alpha\cdot\vec\varphi} -\frac{2}{\lambda}\mu_{-\alpha} \,\right)\,E^{-\alpha}\ .
\label{s2}
\eea
For generic values of the constants $c_1, c_2$ the auxiliary fields are thus fully determined
and can be eliminated from the Lagrangian. 
Let us note that due to the form of the couplings~(\ref{mind}), elimination of the
auxiliary fields $Y$, $Z$ does not introduce any bosonic potential for the hyperscalars
(unlike for the standard YM-hyper couplings~\cite{Sierra:1983uh} where elimination of 
the auxiliary fields introduces a potential
quadratic in the hyper-K\"ahler moment maps $\mu_{ij}$)\,.
The resulting moduli space for the scalars thus is not constrained by any potential.

On the other hand, if $c_2=0$, the auxiliary fields in the Cartan subalgebra remain undetermined 
and we find the constraints $\vec J=0$ and $\vec K= 2\vec\mu/\lambda$. In this case, 
using the expansions (\ref{exp}) with $\vec Y$ and $\vec Z$ as undetermined entries, 
and the remaining components from (\ref{s1}) and (\ref{s2}), and finally setting $c_1=0$ for simplicity, the Lagrangian (\ref{Laux}) becomes
\bea
{\cal L}_Y &=& -\vec Y\cdot \left(\vec K -\frac{2}{\lambda}\vec\mu\right) + \vec Z\cdot \vec J
\nonumber\\[0.5ex]
&& -\frac{2}{|\alpha|^2 \vec\alpha\cdot\vec\varphi} 
\left[J_\alpha\left(K_{-\alpha}-\frac{2}{\lambda} \mu_{-\alpha}\right) - J_{-\alpha}\left(K_{\alpha}-\frac{2}{\lambda} \mu_{\alpha}\right)\right]\ .
\eea
This exhibits the role of the undetermined auxiliary fields as Lagrange multipliers.
In particular the constraint $\vec \mu= \frac{\lambda}2\,\vec K$ modifies the 
hyper-K\"ahler geometry and eliminates degrees of freedom from the hyper-sector.
What we have shown in the above is that such constraints can precisely be avoided 
by introducing abelian factors among the tensor multiplets with the specific couplings~(\ref{mind}).
Let us finally note, that in the Lagrangian~(\ref{SUS6}), the choice of~(\ref{mind}) 
in particular gives rise to interaction terms of the form
\bea
{\cal L}_{\phi{\cal F}^2}&=&  -\tfrac{1}{2}c_1\, \eta_{\ww{m}\ww{n}} \, \phi^0 \,  F_{\mu\nu}{}^{\ww{m}} F^{\mu\nu\,\ww{n}}
 \;,
 \label{c1}
 \eea
and thus to exactly those interactions that were taken into account for the anomaly cancelation 
conditions in~\cite{Blum:1997mm}. 
This led to the selection of $ADE$ gauge groups. Here, we have seen that
such couplings are naturally present in the theory.


\section{Conclusions}


In this paper, we have constructed six-dimensional superconformal models with 
non-abelian tensor and hypermultiplets.
They comprise the field content of $(2,0)$ theories, coupled to $(1,0)$ vector multiplets.
The hypermultiplets are described by gauged nonlinear sigma models with a hyper-K\"ahler 
cone target space and minimal coupling to the superconformal tensor/vector models 
of~\cite{Samtleben:2011fj}. Elimination of the auxiliary fields from the vector multiplets
then further induces non-trivial couplings between hyper and tensor multiplets.
We have shown that proper elimination of the auxiliary fields requires abelian factors
among the tensor multiplets but unlike standard YM-hyper couplings does not give
rise to a scalar potential.
Furthermore, elimination of the auxiliary fields provides couplings (\ref{c1}) that were previously
considered for anomaly cancellations with abelian tensor multiplets and resulted in the selection
of  $ADE$ gauge groups.
We have shown that on the level of the equations of motion, the system may be extended
to include non-propagating three- and four-forms, related by a set of non-abelian first-order duality equations to the physical fields.

It remains an intriguing open question, how much of the presented structures can be carried
over to $(2,0)$ supersymmetric theories. Whereas the tensor and hyper multiplets combine into
the field content of $(2,0)$ tensor multiplets and exhibit some 
unifying structures such as~(\ref{susy_enh}), it is clear that the dynamical degrees of freedom
from the propagating vector multiplets will have to be eliminated upon such a supersymmetry
enhancement. 

The other main open question is of course the quantization of the models, and the fate of the conformal symmetry at the quantum level. For superconformal hypermultiplet actions with and without higher
derivative terms such questions have been addressed in~\cite{Ivanov:2005qf,Ivanov:2005kz}.
For the models presented here at the classical level, a key issue will be whether the
ghost states resulting from the tensor sector of (\ref{SUS6}) can be decoupled with the 
help of the large extended tensor gauge symmetry. This may require to set up a proper
Hamiltonian formalism for the self-dual tensor fields along the 
lines of~\cite{Henneaux:1988gg,Pasti:1996vs}.
Last but not least, the study of anomalies in the generalized gauge symmetries of the models we have presented here will be of great interest.

\subsection*{Acknowledgments}

This work is supported in part by the F\'ed\'eration de Physique Andr\'e Marie Amp\`ere.
R.W. thanks K. Gawedzki and M. Rocek for useful discussions and the Simons Center for Geometry and Physics and the organizers of the Simons Summer Workshop in Mathematics and Physics 2012 for the stimulating environment, as well as the ENS Lyon for hospitality. E.S. would like to thank ENS Lyon for hospitality and Y. Pang and L. Wulff for helpful discussions. The research of E.S. is supported in part by NSF grant PHY-0906222.

\appendix

\section{Conventions}

\label{Aconv}

\paragraph{Indices.} In the main text different kind of indices appear which are
collect in table \ref{tab}. The vector representation of $Sp(1)$ is usually
denoted by an arrow (occasionally indices $\mathsf{i},\mathsf{j} =1,2,3$ are used),
or given in the bi-spinor notation,
\begin{equation}
  \label{A0}
  x^{i}{}_j := \imi\,\vec\sigma^{\, i}{}_j\,\vec x \ ,\quad x^{ij} = \vep^{jk}\, x^i{}_k \quad
  \Rightarrow\quad (x^{ij})^* = \vep_{ik}\,\vep_{j\ell}\, x^{k\ell}=x_{ij}\, ,
\end{equation}
with $\vep^{ik}\,\vep_{jk}=\delta^i_j$ and $\vep^{12}=\vep_{12}=1$.

\begin{table}[!ht]
{\renewcommand{\arraystretch}{1.5}
\renewcommand{\tabcolsep}{0.2cm}
\begin{tabular}{l|c|l}
Label & Range & Comment \\
\hline
$\mu,\, \nu,\ldots$ & $0,\ldots,5$ & world-volume Lorentz indices \\
$\alpha,\, \beta,\ldots$  & $1,\ldots 4n$ & real target space coordinates  \\
$a,\, b,\ldots$ &$1,\ldots, 2n$  & $Sp(n)$ indices, complex coordinates\\
$i,j,\ldots$ &$1,2$ & $Sp(1)$ indices\\
$\hat{\ww{m}},\, \hat{\ww{n}},\ldots$ &$1,\ldots, dim(\hat G)$ & isometries
  of $\ca Q_{4(n-1)}$\\
${\ww{m}},\, {\ww{n}},\ldots$ &$1,\ldots, dim(G)$ & gauged isometries
  of $\ca Q_{4(n-1)}$\\
$I,J,\ldots$ &$1,\ldots,\, n_{\rm T}$ & tensor multiplets\\
$r,s,\ldots$ &$1,\ldots,\, n_{\rm V}$ & vector multiplets\\
\end{tabular}}
\label{tab}
\end{table}

\paragraph{Spinors.}
We work with a flat world-volume metric of signature $(-+++++)$
and Levi-Civita tensor $\varepsilon_{012345}=1$,
$\{\gamma_\mu , \gamma_\nu \} = 2 \eta_{\mu\nu}$ and
$\gamma_7:= \gamma_{012345}$. The spinor chiralities are given by
\begin{equation}
  \label{A1}
  \gamma_7 \,\epsilon^i = \epsilon^i\;,\quad
\gamma_7\, \lambda^{ir} =  \lambda^{ir}\;,\quad
\gamma_7 \,\chi^{iI} = -\chi^{iI}\;,\quad
\gamma_7 \,\psi^a = -\psi^a\, .
\end{equation}
For $Sp(1)$ indices  we use standard
standard northwest-southeast conventions according to
$\epsilon^i = \varepsilon^{ij} \epsilon_j,\, \epsilon_i =  \epsilon^j \varepsilon_{ji}$, etc.\,,
and analogously for the $Sp(n)$ indices, i.e.\ $\psi^a = \Omega^{ab}\,\psi_b,\, \psi_a = \psi^b\,\Omega_{ba}$.
All spinors satisfy a symplectic Majorana condition,
\begin{equation}
\label{A2}
\bar\epsilon_i := \epsilon_i^T C =  (\epsilon^i)^\dagger i \gamma^0\ , \quad
\bar\psi_a :=  \psi_a^T C =  (\psi_a)^\dagger i \gamma^0\, ,
\end{equation}
where the charge conjugation matrix satisfy $\gamma_\mu^T = - C \gamma_\mu C^{-1}$.
Note that indices are exclusively raised/lowered with the symplectic
forms, i.e.\ $ \bar\epsilon^i= - (\epsilon_i)^\dagger i \gamma^0$ etc..
The same relations hold for $\lambda^{ir}$ and $\chi^{iI}$. We refer to the
appendix of \cite{Samtleben:2011fj} for further useful relations.


\section{Target space Geometry}
\label{Ageo}


We collect here some basic relations for the geometrical quantities
of the HK target space.
Many of the following relations are the six-dimensional versions
of the ones given in \cite{deWit:1999fp}.\\

\noindent
{\bf{Vielbeine.}} The $Sp(n)$ vielbeine and connection are
defined in (\ref{vielbein}), such that
metric takes the flat form (\ref{psrctensors}). Thus
for the special case of a flat target space, i.e.\ $g_{\alpha\beta}=\delta_{\alpha\beta}$, the explicit form of the vielbeine is given by ($\kappa^2=i$),
\begin{equation}
  \label{flatVB}
  \begin{bmatrix} f^{1\,a}{}_\alpha \\[5pt] f^{2\,a}{}_\alpha \end{bmatrix} \ =\
   \frac{\kappa}{\sqrt{2}} \begin{bmatrix} \, \Omega & -i\Omega \\
                  -i\unit & \ \unit  \end{bmatrix} \ , \quad
   \begin{bmatrix} f^\alpha{}_{1\,a} & f^\alpha{}_{2\,a} \end{bmatrix} \ =\
   \frac{\bar\kappa}{\sqrt{2}} \begin{bmatrix} \, -\Omega & i\unit \\
                  -i\Omega & \ \unit  \end{bmatrix} \ .
\end{equation}

Besides the metric also the hypercomplex structure and K\"ahler forms (\ref{hk})
can be expressed in terms of the vielbeine:
\begin{equation}
  \label{geo2}
  \omega^{ij}_{\alpha\beta} = 2\, \Omega_{ab}\, f^{(i\,a}{}_\alpha\,f^{j)\,b}{}_\beta\quad , \quad
   \vec J^\alpha{}_\beta  = -i\, f^\alpha{}_{ia}\,\vec\sigma^{\ i}{}_j\, f^{ja}{}_\beta\ ,
\end{equation}
where for the constant flat space vielbeine the first relation reduces to the
expression given in (\ref{psrctensors}). With this the orthogonality of the vielbeine can be written as,
\begin{equation}
  \label{geo0}
 f^{ia}{}_\alpha f^\alpha{}_{jb} = \delta^i{}_j\, \delta^a{}_b\quad , \quad
 f^\alpha{}_{ja}f^{ia}{}_{\beta} = \tfrac{1}{2}\, \left(\, \delta^i{}_j\,  \delta^\alpha{}_\beta
    + i\,\vec\sigma^{\ i}{}_j\, \vec J^{\ \alpha}{}_\beta\, \right)\, ,
\end{equation}
and they satisfy the pseudo reality condition
$(f^{ia}{}_\alpha)^* = \vep_{ij}\,\Omega_{ab}\,f^{jb}{}_\alpha$. With
the definition of the connection (\ref{psrctensors}) this gives,
\begin{equation}
  \label{geo1}
  \omega_\alpha{}^a{}_b = \Omega^{ac}\,\omega_\alpha{}^d{}_c\, \Omega_{db} = - (\omega_\alpha{}^b{}_a)^*
   \quad \Rightarrow \quad
  (\,\Omega\,\omega_\alpha\,)_{[ab]} = 0\, .
\end{equation}
Hence the connection coefficients are $Sp(n)$ matrices (\ref{Xflat}).
\\

\noindent
{\bf{Curvatures.}} The curvatures of the Levi-Civita and $Sp(n)$ connection
are given by
\begin{equation}
  \label{geo3}
  R^\gamma{}_{\delta\alpha\beta} =
   2\,\left(\, \del_{[\alpha}\Gamma^\gamma_{\beta]\delta} + \Gamma^\gamma_{[\alpha| \epsilon}\Gamma^{\epsilon}_{\beta]\delta}\, \right)\;,
  \qquad \ca R^a{}_{b\, \alpha\beta} =
   2\,\left(\, \del_{[\alpha}\omega_{\beta]}{}^a{}_b + \omega_{[\alpha}{}^a{}_c\, \omega_{\beta]}{}^c{}_b\, \right)\,
\end{equation}
The integrability condition of the vielbein postulate (\ref{vielbein}) implies that these
curvature are related in the following way,
\begin{equation}
  \label{geo4}
  R^\delta{}_{\delta\alpha\beta} = f^\gamma{}_{ia}\, f^{ib}{}_\delta\,  \ca R^a{}_{b\, \alpha\beta}\;, \qquad
  \delta^i{}_i \, \ca R^a{}_{b\, \alpha\beta} = f^{ia}{}_\gamma\,f^\delta{}_{jb}\, R^\gamma{}_{\delta\alpha\beta}\, .
\end{equation}
 The symmetries of the Riemann tensor and the first Bianchi identity further imply,
 \begin{equation}
   \label{geo5}
   f^\alpha{}_{ia}f^\beta{}_{jb}f^\gamma{}_{kc}f^\delta{}_{\ell d}\, R_{\alpha\beta\gamma\delta} = \vep_{ij}\, \vep_{kl} \, W_{abcd}\ , \quad
   f^\gamma{}_{kc}f^\delta{}_{\ell d}\, \ca R_{ab\,\gamma\delta} = -\vep_{k\ell}\,  W_{abcd}\, ,
 \end{equation}
where $W_{abcd} =  W_{(abcd)}$ is the totally symmetric curvature tensor,
with reality property $(W_{abcd})^* = W^{abcd}$, that appears in the
Lagrangian (\ref{La}). $Sp(n)$ indices are raised/lowered as described in appendix~\ref{Aconv}.
From the second Bianchi identity it follows that
\begin{equation}
  \label{geo6}
  \nabla_{[\alpha}R_{\beta\gamma]\delta\epsilon} = 0 \quad \Rightarrow \quad f^\alpha{}_{ia}\, \ca D_\alpha W_{bcde} =
  f^\alpha{}_{i(a}\, \ca D_\alpha W_{bcde)}\, ,
\end{equation}
where $\ca D_\alpha$ is the covariant derivative w.r.t.\ the $Sp(n)$ connection.
\\

\noindent
{\bf{Isometries.}} The vielbeine introduced here are adjusted to the
$\mathrm{H}_{Sp(1)}\otimes \mathrm{P}_{Sp(n)}$ structure of the tangent space (\ref{vielbein}).
They therefore relate the coordinate basis to vector fields of the form,
\begin{equation}
  \label{geo7}
  e_{ia} = \theta_i\otimes e_a = f^\alpha{}_{ia}\del_\alpha\, ,
\end{equation}
with an analogous relation for the dual basis with the inverse vielbein.
The Lie derivative of these vector fields along  a vector field $X$ is then given by,
\begin{equation}
  \label{geo8}
  \mathscr{L}_X\, e_{ia} = \left[\, \delta^j{}_i (\,X^\alpha\,\omega_\alpha{}^b{}_b -t^b{}_a\,) -
     \tfrac{1}{2}\,\vec\sigma^{\ j}{}_i\, \vec t^{\ b}{}_a\, \right] e_{jb}\ ,
   \quad  t^{\ b}{}_a = \tfrac{1}{2}\, f^{ib}{}_\alpha(\nabla_\beta X^\alpha)f^\beta{}_{ia}\, ,
\end{equation}
where $t^{\ b}{}_a$ was introduced below (\ref{dtot}) and
$\vec t^{\ b}{}_a = f^\beta{}_a\,\vec\sigma\, f^{b}{}_\alpha (\nabla_\beta X^\alpha)$. For
diffeomorphism or isometries that commute $Sp(1)$ isometries (\ref{hkp}), and thus preserve the
$\mathrm{H}_{Sp(1)}\otimes \mathrm{P}_{Sp(n)}$ structure, the latter matrices vanish. In that
case one has,
\begin{equation}
  \label{geo9}
 \mathscr{L}_X\, e_{a} =  (\,X^\alpha\,\omega_\alpha{}^b{}_b -t^b{}_a\,)\, e_{b}\quad \Rightarrow\quad
 \mathscr{L}_X\, W_a = X^\alpha\, \ca D_\alpha W_a  + t^b{}_a W_b\quad  \mathrm{etc.}
\end{equation}
In the case that $X$ is also an isometry, $\nabla_{(\alpha}X_{\beta)} = 0$ one finds,
\begin{equation}
  \label{eq:2}
  \Omega^{ac}\,t^d{}_c\, \Omega_{db} = t^a{}_b  = - (t^b{}_a)^*
  \quad  \Rightarrow \quad (\,\Omega\, t\,)_{[ab]} = 0,
\end{equation}
These relations are the same as for the $Sp(n)$ connection (\ref{geo1}). In the case that
the isometries $X_{(\ww{m})}$ obey
$[\, X_{(\ww{m})}\, ,\, X_{(\ww{n})}] = -f_{\ww{m}\ww{n}}{}^{\ww{p}}\,  X_{(\ww{p})}$ one also finds,
\begin{equation}
  \label{geo10}
  \ca D_\alpha\, t_{(\ww{m})}{}^a{}_b = \ca R^a{}_{b\, \alpha\beta}\, X_{(\ww{m})}^\beta \ ,
  \quad [\, t_{(\ww{m})}\, ,\, t_{(\ww{n})}\, ]^a{}_b
    = f_{\ww{m}\ww{n}}{}^{\ww{p}} \, t_{(\ww{p})}{}^a{}_b
    +   \ca R^a{}_{b\, \alpha\beta}\, X^\alpha_{(\ww{m})} X^\beta_{(\ww{n})} \, .
\end{equation}

We finally mention the equivariance condition for the moment maps of triholomorphic
isometries $X_{(\ww{m})}$ (\ref{mmap}). The identity
$i_{[X_{(\ww{m})},X_{(\ww{n})}]}\, \vec\omega =
 [\, \mathscr{L}_{X_{(\ww{m})}}, i_{X_{(\ww{n})}}\, ]\,\vec\omega $ implies for triholomorphic
isometries,
\begin{equation}
  \label{ge11}
  \mathscr{L}_{X_{(\ww{m})}} \vec\mu_{\ww{n}} =
   \vec \omega_{\alpha\beta}X^\alpha_{(\ww{m})}X^\beta_{(\ww{n})} = -f_{\ww{m}\ww{n}}{}^{\ww{p}}\vec\mu_{\ww{p}}
   + const.\, ,
\end{equation}
where superconformal symmetry fixes the constant to be zero.

\bigskip
\bigskip
\bigskip




\providecommand{\href}[2]{#2}\begingroup\raggedright\endgroup

\end{document}